\documentclass{moriond}

\def\be{\begin{equation}}
\def\ee{\end{equation}}
\def\bea{\begin{eqnarray}}
\def\eea{\end{eqnarray}}

\usepackage{graphics,epsfig,color} 
\usepackage{varioref,exscale,latexsym,amsmath,amssymb}
\usepackage{hyperref}
\usepackage{slashed}
\def\d {\mbox{d}}
\def\L {\mathcal{L}}

\newcommand{\w}[1]{\tilde{#1}}

\def\be{\begin{equation}}
\def\ee{\end{equation}}
\def\bea{\begin{eqnarray}}
\def\eea{\end{eqnarray}}

\newcommand{\Photo}{}

\begin{document}
\vspace*{4cm}
\title{BASICS OF THE PRESSURON}

\author{O. MINAZZOLI$^{1}$, A. HEES$^2$}

\address{$^{1}$ Centre Scientifique de Monaco, 8 Quai Antoine 1er, Monaco\\
UMR ARTEMIS, CNRS, University of Nice Sophia-Antipolis,
Observatoire de la C\^ote d’Azur, BP4229, 06304, Nice Cedex 4, France\\
$^2$Department of Mathematics, Rhodes University, Grahamstown 6140, South Africa}

\maketitle\abstracts{
The pressuron is a specific case of a dilaton-like field that leads to a decoupling of the scalar-field in the field equation for pressureless fluids. Hence, the pressuron recovers general relativity in the limit of weak pressure. Here we review its basics.}

\section{Introduction} 

Perturbative string theory predicts a scalar partner to the metric called dilaton. At tree-level, the dilaton couples multiplicatively to both the Ricci scalar and matter, and is massless \cite{damour:1994fk}. Hence, the dilaton should lead to a  violation of the various flavours of the Einstein equivalence principle. But for the theory to be potentially valid, it must be able to satisfy the strong observational constraints existing on the equivalence principle. Unfortunately, pertubative string theory cannot predict the effective form of the dilaton couplings and therefore cannot explain how the dilaton is supposed to satisfy those constraints. A handful set of (bottom-up) solutions have been proposed. The pressuron is one of them --- although its phenomenology could be studied independently of its speculative fundamental origin.

\section{Decoupling in pressurless regimes}
\label{sec:pressurless}
 
Assuming that one can factorize the scalar field contribution to the mass out --- which can be either an approximation (see section \ref{sec:real}), or come from a more fundamental (top-down) reason that would have to be understood \footnote{Note for instance that in Bars, Steinhardt and Turok \cite{bars:2014pr}, local conformal symmetry in the particle sector implies that all masses are proportional to the non-minimally coupled scalar-field $h$ --- that is identified to the Higgs field in their fully Weyl invariant model of fundamental physics (ie. gravitation and particle physics).} --- the action of the pressuron in the string frame can be written generally as

\be
S= \frac{1}{c}\int  d^4x \sqrt{-g} \frac{1}{2\kappa}\left(h^2  R-Z(h) (\partial h)^2 - V(h) \right)   -c^2 \sum_A \int h m_A d\tau_A, ~~~~\textrm{ with } \frac{\d m_A}{\d \tau_A}=0\label{eq:actionhiggsNMC},
\ee
where $\kappa$ is the gravitational coupling constant, $g$ is the determinant of the space-time metric $g_{\mu\nu}$, $h$ is a scalar field, $Z(h)$ and $V(h)$ are arbitrary functions. Note that in this model all atom masses as well as the Planck mass are proportional to the pressuron field. We consider both massless and self-interacting cases to stay as general as possible. Considering the standard model of particles at low energies and in the chiral limit, one would have $m_A \sim \textrm{A} N \Lambda$, where $\Lambda$ is the quantum chromodynamics (QCD) renormalisation group invariant mass scale, $N$ a pure number and A the atomic (mass) number of the considered atom \cite{gasser:1982ph}. The material part of the Lagrangian can be rewritten equivalently as follows:
\bea
S_m&=&-\frac{1}{c}\int  d^4x \sqrt{-g} h c^2\rho,\textrm{ with } \nabla_\sigma (\rho U^\sigma)=0, \label{eq:effectLpl}\\
 \rho&=&\sum_A (\sqrt{-g} U^0)^{-1} m_A \delta^{(3)}(x^\alpha-x_A^\alpha), \textrm{ and } U_A^\alpha \equiv \frac{dx^\alpha}{c d \tau_A}, \nonumber
\eea
where $\delta^{(3)}$ is the 3-Dirac delta function. It is obvious that one can rewrite action in Eq. \ref{eq:actionhiggsNMC} in the Einstein frame where the scalar field no-longer couples to neither the metric nor to the material fields
\be
S= \frac{1}{c}\int  d^4x \sqrt{-\w{g}} \frac{1}{2\kappa}\left( \w{R}-\w{Z}(h) (\w{\partial} h)^2 - \w{V}(h) \right)   -c^2 \sum_A \int  m_A d\w{\tau}_A, ~~~~\textrm{ with } \w{g}_{\mu \nu} = h^2 g_{\mu \nu}.
\ee
Therefore, as long as dust fields are considered, the theory is not different from general relativity. In other words, in the pressureless limit, the pressuron theory is nothing but a ``Veiled'' general relativity \cite{deruelle:2011nd}. The question now is: what happens when material fields are pressureful?

\section{From pressurless to pressureful fluids}
\label{sec:pressure}

One searches the Lagrangian $\L_m$ suitable for pressureful perfect fluids --- where one uses the following definition $S_m = 1/c~\int   d^4x \sqrt{-g} h \L_m$. In the pressureless case, Eq.~\ref{eq:effectLpl} shows that $\L_m=- c^2 \rho$ . One can show that the conservation equation $\nabla_\sigma (\rho U^\sigma)=0$ --- coming from the assumption that the pressuron field contribution to the mass can be factorized out (ie. $dm_A/d\tau_A=0$ in Eq.~\ref{eq:actionhiggsNMC}) --- induces a strict relation between the rest mass energy density variation and the metric field variation given by \cite{harko:2010pd} $\delta \rho = 1/2~ \rho(g_{\mu \nu}+U_\mu U_\nu) \delta g^{\mu \nu}$. Using this relation, one can derive the following equation for the stress-energy tensor of a barotropic fluid:
\be
T_{\mu \nu} \equiv \frac{2}{\sqrt{-g}} \frac{\partial (\sqrt{-g} \L_m)}{\partial g^{\mu \nu}}=- \rho \frac{d \L_m}{d \rho} U_\mu U_\nu +\left(\L_m -  \rho \frac{d \L_m}{d \rho} \right)g_{\mu \nu}.
 \ee
Since we want the Lagrangian $\L_m$ of a perfect fluid, one simply has to equalize this equation to the stress-energy tensor of a perfect fluid $T_{\mu \nu}=(\epsilon+P) U_\mu U_\nu + P g_{\mu \nu}$, where $\epsilon$ is the total energy density of the fluid and $P$ its pressure. This leads to a set of two first order linear equations whose solution when considering a barotropic fluid is \cite{minazzoli:2012pr} $\L_m=-\epsilon(\rho)$. Therefore, the action considered in Eq.~\ref{eq:actionhiggsNMC} for a barotropic perfect fluid becomes \cite{minazzoli:2013fk}
\be
S= \frac{1}{c}\int  d^4x \sqrt{-g} \left[ \frac{1}{2\kappa}\left(h^2  R-Z(h) (\partial_\sigma h)^2 - V(h) \right)   - h \epsilon\right], ~~~~\textrm{ with } \epsilon(\rho) = c^2 \rho + \rho \int \frac{P(\rho)}{\rho^2} d \rho \label{eq:actionhiggsNMC2}. 
\ee
This way, one recovers the usual conservation equation for the total energy density $\nabla_\sigma(\epsilon U^\sigma)=-P \nabla_\sigma U^\sigma$. Note however that the equation of motion is unconventional: diffeomorphism invariance of action in Eq. (\ref{eq:actionhiggsNMC2}) induces $\nabla_\sigma T^{\mu \sigma}=-(\epsilon g^{\mu \sigma}+T^{\mu \sigma}) \partial_\sigma \ln h$. Now, one can check that in comparison to the pressureless case, the scalar field in the Einstein frame action is no longer totally decoupled from matter because of a remaining pressure term. Hence the theory is no longer equivalent to general relativity when there is pressure. In some sense, pressure breaks the equivalence between the pressuron action and general relativity. Indeed, in both Einstein and string frames, the pressuron is sourced by pressure and not by energy density  (i.e. roughly speaking, one has $\Box h \propto P$ instead of $\Box h \propto \epsilon$) \cite{minazzoli:2013fk,minazzoli:2014ao,minazzoli:2014pb}. For instance, in the string frame, the scalar field equation reads
\be
\Box h +\frac{1}{h} \left[1+\frac{h}{2}\frac{Z_{,h}(h)}{Z(h)+6} \right] (\partial h)^2 = \frac{\kappa ~ 3  P}{Z(h)+6} +  \frac{V_{,h}(h)/2-2V(h)/h}{Z(h)+6}.
\ee
Fortunately, in weak gravitational fields such as in the solar system, bodies pressure is orders of magnitude lower than their energy density (e.g. $P/(c^2\rho) \sim 10^{-6}$ for the Earth). Therefore, effects of the pressuron field are drastically reduced in low pressure regions such as in the solar system \cite{minazzoli:2013fk} or during the cosmological matter era \cite{minazzoli:2014ao,minazzoli:2014pb}, whether the pressuron is massless or not. Hence, the pressuron is not constrained by current solar system gravitational tests as long as $Z$ is not close to the singular value $Z=-6$. But as in usual Brans-Dicke theory, the singularity corresponds to an infinite coupling function in the Einstein frame such that it cannot be reached dynamically \cite{minazzoli:2014ao,minazzoli:2014pb}. Note also that $Z$ can be re-written in terms of the usual Brans-Dicke omega function via \cite{minazzoli:2013fk} $Z(h) = 4 ~\omega(h^2)$. On the other side, depending on the coupling function $Z(h)$, one can expect to see a pressuron's signature in strong field regimes, where pressure cannot be neglected. Hence, testing the theory in these regimes may be the only way to reasonably constrain it. In particular, it would be interesting to see whether or not a spontaneous scalarization can occur as in standard scalar-tensor theories for specific parameters \cite{damour:1996uq}.

\section{Cosmology}

During the matter era the matter content of the universe is dominated by pressure-less fields. Therefore, the massless theory  quickly converges towards a constant scalar field regardless the pressuron's value at the transition between the radiation and the matter era \cite{minazzoli:2014ao,minazzoli:2014pb}. For that reason, the pressuron cannot explain dark energy by itself and needs either a cosmological constant or a self-interaction potential --- or some feedback effects --- in order to explain the apparent acceleration of the expansion of the universe \cite{minazzoli:2014ao}. This may be seen as a drawback of the theory only if one believes that dilaton fields should also play the role of dark energy. However, there is no fundamental reason to believe so at the moment. Also because of the quick convergence of the pressuron during the matter era, the electromagnetic-pressuron coupling is not constrained by cosmological observations \cite{hees:2014pr,hees:2015gg} at low redshift.

The radiation era is more subtle when there is a multiplicative scalar-matter coupling because it is not obvious if the perfect fluid on-shell Lagrangian $\L_m= - \epsilon$ can still be used in this era. This subject is currently under investigation. However, it is expected that unlike in ``standard'' scalar-tensor theories \cite{damour:1993kx}, the reduced scalar-field equation in the Einstein frame will keep a potential term during the radiation era, leading to a Damour and Nordtvedt dynamical decoupling mechanism  \cite{damour:1993kx} during this period for a subclass of functions \cite{minazzoli:2014pb} $Z$. In this scenario, $Z$ would be dynamically driven toward a big value during the radiation era. But unlike in standard scalar-tensor theory, this mechanism is not mandatory in order to explain solar system observations since the pressuron is automatically already weakly coupled in regions with low pressure.

\section{From universality to a more likely picture}
\label{sec:real}

In section \ref{sec:pressurless}, one assumes that the dilaton contribution can be factorized out of the particle masses at the level of the effective hadronic action. However, due to the complicated nature of the various contributions to the mass of nucleons, a more likely picture would be that the dilaton field cannot be fully factorized out, but that there would remain some functional dependencies of the masses with respect to the dilaton. In the action of Eq.~\ref{eq:actionhiggsNMC}, it would mean that one effectively has $m_A = m_A(h)$ and therefore $dm_A/d\tau \neq 0$. 

However, since most of nucleons mass comes from the gluonic interaction, it is sufficient to suppose that the dilaton couples multiplicatively to the chiral limit of nucleons mass (ie. $\L_g \propto h \Lambda$, where $\Lambda$ is the renormalisation group invariant QCD mass scale) to get a partial decoupling characterized by the fact that the main microscopic contribution to the particles mass (but not all contributions) will cancell out in the scalar-field equation. In that picture, one would get deviations from general relativity, but weaker than for general dilaton fields. In addition, a linear coupling to the fermions mass as well can enhance this decoupling. In this optimistic scenario, only a few 0.1\% of nucleons mass corresponding to the photons cloud could contribute to terms that violate the equivalence principle. In that case, one expects to get equivalence principle violating terms similar to those computed in Damour and Donoghue \cite{damour:2010zr}, but with a decrease of about four orders of magnitude compared to the general dilaton case. A derivation of the actual numerical decoupling amplitudes in the chiral perturbation theory framework for various set of dilatonic parameters is on its way.  Also, in this picture, one shall have to re-derive the fluid limit with $dm/d\tau \neq 0$. Indeed, let us recall that $\L_m=-\epsilon$ is solution of a set of equations that one expects to be modified when $dm/d\tau \neq 0$. Hence, it is likely that one will get a slightly different effective fluid Lagrangian, that nonetheless has to reduce to $\L_m = - \epsilon$ in the universal limit considered in sections \ref{sec:pressurless} and \ref{sec:pressure} (ie. $d m/d\tau \rightarrow 0$). The question relative to the effective fluid description in this more likely picture is also under consideration.

\section{Conclusion}

It is often believed that a massless scalar field that would couple to both the Ricci and matter in the action necessarily leads to strong departures from what is observed in our solar system. However, we found a specific phenomenological example for which it is not the case. From a theoretical point of view, it would be interesting to see if one could imagine a top-down justification to the specific effective pressuron coupling, may it be in the string framework or not. From a phenomenological point of view, whether it is massless or not, one should try to derive the full pressuron phenomenology in order to find observational ways to constrain it.

\section*{Acknowledgments}

The authors are really grateful to the organizers for having them get financial support to attend the conference. In addition, AH acknowledges support from ``Fonds Sp\'ecial de Recherche" through a FSR-UCL grant.


\section*{References}

\bibliography{biblio_COPY} 
%
%
%
%
%
%
%
%
%
%
%
%
%
%


\end{document}